\documentclass{desyprocA4}
\def\eslt{E_T^{\rm miss}}

\def\to{\rightarrow}

\def\bi{\begin{itemize}}
 \def\ei{\end{itemize}}

\def\c1p{C1^\prime}
\def\msq3{\overline{m}_{\tilde{q}}(3)}

\def\tg{\tilde g}

\def\tq{\tilde q}

\def\be{\begin{equation}}  
\def\ee{\end{equation}}  
\def\bea{\begin{eqnarray}}  
\def\eea{\end{eqnarray}}  
\def\tw{\widetilde W}

\def\tz{\tilde Z}

\def\beq{\begin{equation}}
\def\eeq#1{\label{#1}\end{equation}}
\def\eeqn{\end{equation}}

\newenvironment{Eqnarray}%
   {\arraycolsep 0.14em\begin{eqnarray}}{\end{eqnarray}}
\def\beqa{\begin{Eqnarray}}
\def\eeqa#1{\label{#1}\end{Eqnarray}}
\def\eeqan{\end{Eqnarray}}

\begin{document}
\title{SUSY discovery potential of LHC14 with 0.3-3~ab$^{-1}$ :\\ A Snowmass
  whitepaper.}

\author{{\slshape H. Baer$^1$, V. Barger$^2$, A.~Lessa$^3$ and X. Tata$^4$},\\
  $^1$Dept. of Physics and Astronomy, University of Oklahoma, Norman, OK 73019, USA\\ 
$^2$  Dept. of Physics, University of Wisconsin, Madison, WI 53706, USA\\
$^3$Instituto de F\'isica, Universidade de S\~ao Paulo, S\~ao Paulo-SP, Brazil\\
$^4$Dept. of Physics and Astronomy, University of Hawaii at Manoa,
    Honolulu, HI 96822, USA}
%


\maketitle


\begin{abstract}
We examine the discovery reach of LHC14 for supersymmetry for integrated luminosity
ranging from 0.3 to 3~ab$^{-1}$. 
In models with gaugino mass unification and $M_1,\ M_2\ll |\mu|$ (as for mSUGRA/CMSSM),
we find a reach of LHC14 with 3~ab$^{-1}$ for gluino pair production 
extends to $m_{\tg}\sim 2.3$ TeV while the reach 
via $\tw_1\tz_2 \to Wh+\eslt$ extends to $m_{\tg}\sim 2.6$ TeV.
\end{abstract}

%
Recently, the European Strategy for Particle Physics group has commissioned
studies on the discovery potential of high luminosity options of LHC
operating at $\sqrt{s}\simeq 14$ TeV\cite{Akesson:2006we}. 
Integrated luminosity values ranging from 0.3-3 ab$^{-1}$ have been 
considered\cite{ATLAS-Collaboration:2012jwa}.

To assist this program, we presented computations in Ref. \cite{reach} 
of the high luminosity reach of LHC14 
for discovery of supersymmetry within the context of the popular mSUGRA/CMSSM 
model (although our results should be valid more generally for most SUSY models 
with gaugino mass unification and $M_1,\ M_2\ll |\mu|$). 
We examined the SUSY reach via the usually considered 
gluino and squark pair production reactions as well as from electroweak gaugino production. 
For very high integrated luminosities at the ab$^{-1}$ range, the gaugino pair
production reactions offers a larger reach opportunity since at very high mass values
gluino and squark production becomes kinematically suppressed.  
We present our results here in an abbreviated summary form 
as a contribution to the US Snowmass Energy Frontier planning process.

We begin by considering the multi-jet + multi-lepton + $\eslt$ signal
that arises from gluino and squark pair production, followed by their
cascade decays to charginos and neutralinos, with the decay chain
terminating in a stable LSP that is the origin of $\eslt$. 
Following Ref. \cite{reach}, we classify the events by lepton multiplicity, with
additional requirements on jets:
\bi
\item $0l$: $n(l)=0$, $n(j) \ge 3$, $\{E_T(j_1),E_T(j_2),E_T(j_3)\} >
$\{100 GeV, 100 GeV, 50 GeV\};
\item $1l$: $n(l)=1$, $n(j) \ge 2$, $\{E_T(j_1),E_T(j_2)\} > $\{100 GeV,
100 GeV\};
\item $2l$: $n(l)=2$, $n(j) \ge 2$, $\{E_T(j_1),E_T(j_2)\} > $\{300 GeV,
300 GeV\}.  
\ei 
We also evaluate dominant SM backgrounds to these topologies from
$t\bar{t}$, $W$+jets, $Z \to \ell\ell$+jets, $Z\to \nu\nu$+jets and
$Zt\bar{t}$ production. We deem the signal to be observable over the
background after a $\eslt > \eslt({\rm min})$ cut if the number of
signal events exceeds $max\left[\ 5\ {\rm events},\ 0.2N_B,\
5\sqrt{N_B}\right]$ for a specified value of  the integrated luminosity. Here
$N_B$ equals the corresponding number of background events.  We optimize
the signal relative to background by varying $\eslt({\rm min})$ between
100-1500~GeV in 100~GeV steps.

\begin{figure}[htb]
  \begin{center}
\includegraphics[width=0.9\textwidth]{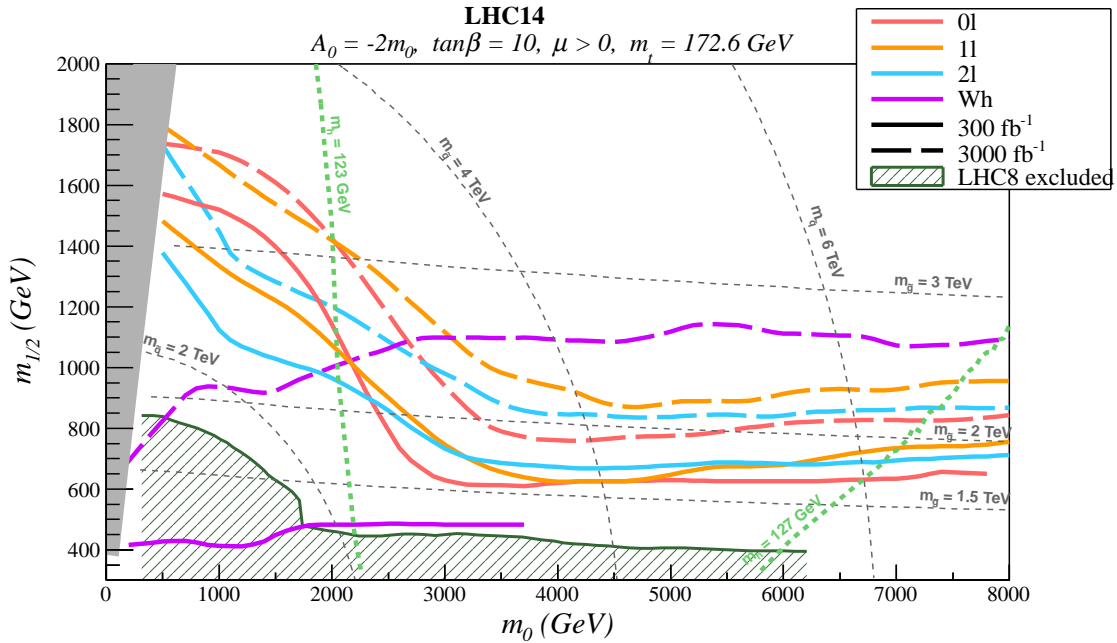}
  \end{center}
\caption{SUSY reach in the various channels discussed in the text for LHC14
 for integrated luminosities of 300 fb$^{-1}$ (solid lines) and 3000
 fb$^{-1}$ (dashed lines).  
The shaded  grey area on the left side of the figure is excluded 
because the stau
 becomes the LSP. The green shaded region in lower-left and extending
 across the bottom is excluded 
 by SUSY searches at LHC8 \cite{lhc8}.}
\label{fig:reach}
\end{figure}

The LHC reach in each of these channels\footnote{We have deliberately
 not shown results for the rate-limited low background
 same sign dilepton and trilepton channels because we were unable to
 reliably estimate the backgrounds for these very high integrated
 luminosity values. Also hard-to-estimate backgrounds from lepton fakes or
 charge misidentification could also make substantial contributions in
 these channels.}
 is presented in Fig.~\ref{fig:reach} where we show the $m_0-m_{1/2}$ plane for
$\tan\beta =10$ and $A_0=-2m_0$. The large $A_0$ value is necessary to allow for large mixing 
in the top-squark sector, which is required to accommodate a Higgs mass 
$m_h \sim 125$~GeV. 
The solid (dashed) lines are for an integrated luminosity of 300 (3000)~fb$^{-1}$. 

For very large $m_{1/2}$, gluino and squark pair production cross-sections are 
suppressed in part by low PDF luminosities at large $\hat{s}$. 
In this case, the wino pair production reactions $pp \to \tw_1\tw_1$ or $\tz_2\tw_1$ 
become the dominant SUSY production processes, even more so in the 
case where squarks are also heavy\cite{wss}, {\it i.e.} large $m_0$. 
For the wino pair production reaction, the chargino typically decays via $\tw_1\to W\tz_1$ while
the neutralino decays via $\tz_2 \to h\tz_1$ if $m_{\tz_2}-m_{\tz_1} > m_h$. 
The signals from chargino pair production are typically buried below SM
backgrounds from $WW$, $Wj$ and $t\bar{t}$ production. 
Following Ref.~\cite{warin}, we focus on the $\tw_1\tz_2\to Wh+\eslt\to \ell b\bar{b}+\eslt$ 
signal. 
To extract signal from various backgrounds, we require 
\bi
\item $n(l)=1$, $n(b)= n(j)= 2$, $\Delta \phi(b,b) < \pi/2$, $M_{eff} >
350$ GeV, $m_{T}(\ell,\eslt) > 125$ GeV, 100 GeV $< m_{bb} < $ 130 GeV.
\ei 
Here $M_{eff}=\sum_{i} E_T(j_i) + \sum_{i} p_T(l_i) + \eslt$,
$m_T(\ell,\eslt)$ is the transverse mass and $m_{bb}$ the invariant mass
of the b-jet pair. As before, we optimize with respect to $\eslt({\rm
min})$.  The reach via this $Wh$ channel is shown by the purple curves
in Fig.~\ref{fig:reach}. We see that while the strong production
dominates the LHC reach for 300~fb$^{-1}$, {\it the reach via the $Wh$
channel exceeds that from gluino production (if squarks are heavy) for
an integrated luminosity of 3~ab$^{-1}$.}

Our results for the reach, expressed in terms of $m_{\tg}$, are summarized in Table~\ref{tab:reach}.
\begin{table}
\centering
\begin{tabular}{|c|c|c|c|}
\hline
IL (fb$^{-1}$) &  $m_{\tq}\sim m_{\tg}$ & $m_{\tq}\gg m_{\tg}$ & $Wh$   \\
\hline
100  &  3.0 TeV  & 1.6 TeV  & - TeV \\
300  &  3.2 TeV  & 1.8 TeV  & 1.2 TeV \\
1000 &  3.4 TeV  & 2.0 TeV  & 2.0 TeV \\
3000 &  3.6 TeV  & 2.3 TeV  & 2.6 TeV \\
\hline
\end{tabular}
\caption[]{Optimized SUSY reach of LHC14 within the mSUGRA/CMSSM model
expressed in terms of $m_{\tg}$ for various choices of integrated
luminosity.  
The $m_{\tq}\sim m_{\tg}$ and $m_{\tq}\gg m_{\tg}$ values
correspond to the maximum reach in the $0l$, $1l$ and $2l$
channels from gluino and squark pair production 
while the $Wh$ values shown correspond to the reach in the $Wh$ channel for
$m_{\tq}\gg m_{\tg}$.  }
\label{tab:reach}
\end{table}
Although we have illustrated the results using the mSUGRA/CMSSM framework, we
expect that the qualitative features of the Table will be valid in any
model with gaugino mass unification and large $|\mu|$. 
In contrast, in models where $|\mu| \ll |M_{1,2}|$ , then 
wino pair production leads to a striking hadronically-quiet same-sign diboson
signal with $\ell^\pm\ell^\pm+\eslt$ final state that again yields a
larger reach than gluino and squark pair production for integrated luminosities
greater than 300~fb$^{-1}$ \cite{lhcltr}.

\section*{Acknowledgments} This research was sponsored in part by grants
from the US Department of Energy and FAPESP.


\begin{footnotesize}

\end{footnotesize}


\end{document}